\documentclass[a4paper,12pt]{article}
\pdfoutput=1 
\usepackage{jcappub-mod}
\usepackage[T1]{fontenc}
\usepackage[normalem]{ulem}
	
\title{\textbf{Non-perturbative results for the \\
luminosity and area distances}}

\author[a,b]{Dimitar Ivanov,}
\author[a,b]{Stefano Liberati,}
\author[a,b,c]{Matteo Viel,}
\author[d]{\\and Matt Visser\,}

\affiliation[a]{SISSA - International School for Advanced Studies, \\
                   via Bonomea 265, 34136 Trieste, Italy.}

\affiliation[b]{INFN, sezione di Trieste, via Valerio 2, Trieste, Italy.}

\affiliation[c]{INAF, Osservatorio Astronomico di Trieste, \\
                    via Tiepolo 11, I-34131 Trieste, Italy.}

\affiliation[d]{School of Mathematics and Statistics,
Victoria University of Wellington;\\
PO Box 600, Wellington 6140, New Zealand.}

\emailAdd{divanov@sissa.it}
\emailAdd{liberati@sissa.it} 
\emailAdd{viel@sissa.it}
\emailAdd{matt.visser@sms.vuw.ac.nz}

\abstract{
The notion of luminosity distance is most often defined in purely FLRW (Friedmann--Lemaitre--Robertson--Walker) cosmological spacetimes, 
or small perturbations thereof. However, the abstract notion of luminosity distance is actually much more robust than this, and can be defined non-perturbatively in almost arbitrary spacetimes. Some quite general results are already known, in terms of $dA_\mathrm{observer}/d\Omega_\mathrm{source}$,  the cross-sectional area per unit solid angle of a null geodesic spray emitted from some source and subsequently detected by some observer. We shall reformulate these results in terms of a suitably normalized null geodesic affine parameter and the van Vleck determinant, $\Delta_{vV}$.  The contribution due to the null geodesic affine parameter is effectively the inverse square law for luminosity, and the van Vleck determinant can be viewed as providing a measure of deviations from the inverse square law. This formulation is closely related to the so-called Jacobi determinant, but the van Vleck determinant has somewhat nicer analytic properties and wider and deeper theoretical base in the general relativity, quantum physics, and quantum field theory communities. In the current article we shall concentrate on non-perturbative results, leaving near-FLRW perturbative investigation for future work. 

\medskip\noindent
\emph{Date:} 19 February 2018; 26 February 2018; 30 April 2018; LaTeX-ed \today.

}
\begin{document}
\maketitle
\flushbottom
\def\d{{\mathrm{d}}}
\def\g{{\mathfrak{g}}}
\def\dbar{{\mathchar'26\mkern-12mu \d}} 
\def\tr{{\mathrm{tr}}}
\def\Hilbert{{\mathcal{H}}}
\def\H{{\mathrm{H}}}
\def\R{{\mathrm{R}}}
\def\E{{\mathrm{E}}}
\def\arcsinh{{\mathrm{arcsinh}}}
\def\lint{\hbox{\Large $\displaystyle\int$}}   
\def\hint{\hbox{\huge $\displaystyle\int$}} 
\def\tr{{\mathrm{tr}}}
\definecolor{purple}{rgb}{1,0,1}
\newcommand{\red}[1]{{\slshape\color{red} #1}}
\newcommand{\blue}[1]{{\slshape\color{blue} #1}}
\newcommand{\purple}[1]{{\slshape\color{purple} #1}}
\parindent0pt
\parskip7pt
\def\pdet{{\mathrm{pdet}}}
\section{Introduction}\label{S:intro}

\begin{figure}[h!]
\centering
\includegraphics[width=0.5\textwidth]{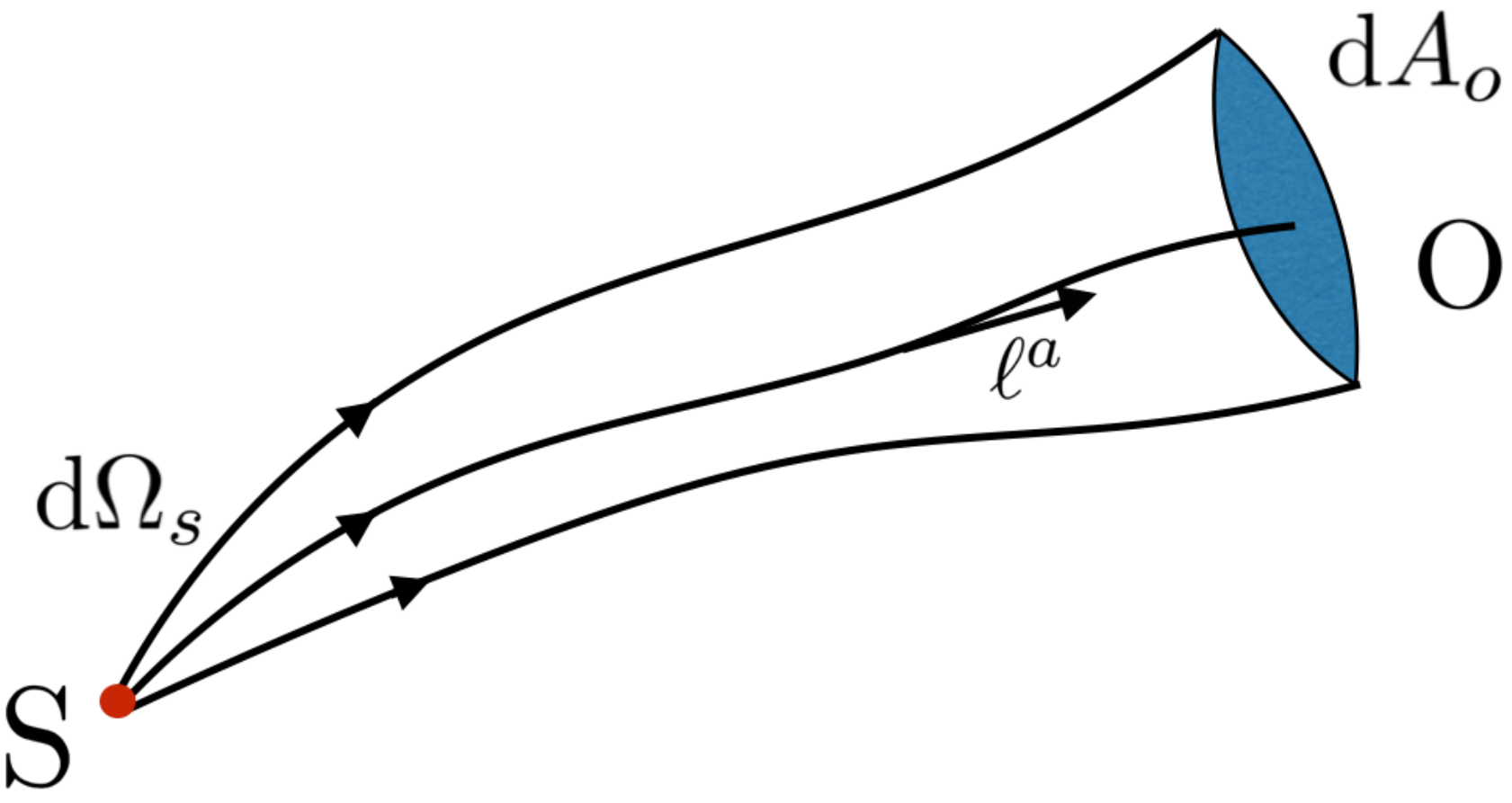}
\caption{\label{F:extended-observer}
A congruence of light rays emitted at a point source $S$ and received by an extended observer $O$. The luminosity distance between $S$ and $O$ is given by $d_L = (1+z)\sqrt{\mathrm{d}A_o/\mathrm{d}\Omega_s}$,
while the area distance is  $d_{area} = \sqrt{\mathrm{d}A_o/\mathrm{d}\Omega_s}$. Here $\ell^{a}$ is a tangent vector to a geodesic in the congruence. }
\end{figure}

The luminosity distance, or its variant the luminosity modulus, is one of the key semi-empirical observational quantities used in cosmology~\cite{Bonvin:2005,Sasaki:1987,Futamase:1989,Pyne:2003,Barausse:2005,Hui:2005,Fanizza:2013,Marozzi:2014,Yoo:2016,Umeh:2012,Umeh:2014,DiDio:2016} and cosmography~\cite{Visser:2003,Visser:2004,Cattoen:2007a,Cattoen:2007b,Cattoen:2008,Visser:2009,Vitagliano:2009,Visser:2015,Visser:2015b}.
Despite its widespread use and popularity, the theoretical foundations of the notion of luminosity distance still leave a number of open issues, which we shall explore in the current manuscript.
In particular, in the current article we will focus on the extent to which we can make non-perturbative statements in generic spacetimes and we shall show that the luminosity distance is non-perturbatively related to total redshift, suitably normalized null affine parameter, and van Vleck determinant $\Delta_{vV}$ by
\begin{equation}
d_L = (1+z)\; \sqrt{dA_o\over d\Omega_s} =  (1+z) \; (\lambda_o-\lambda_s) \; \Delta_{vV}^{-1/2}.
\end{equation}
We also address the closely related ``area distance'', (see figure \ref{F:extended-observer}), physically related to counting the number of photons in a burst,  and demonstrate
\begin{equation}
d_{area} = \sqrt{dA_o\over d\Omega_s} =  (\lambda_o-\lambda_s) \; \Delta_{vV}^{-1/2}.
\end{equation}
A perturbative analysis,  addressing  ``close to FRLW'' cosmologies,  will be deferred for future work~\cite{future}.

There are other approaches to deriving non-perturbative results for the luminosity distance in the literature. One possibility is to consider special adapted coordinate systems which greatly simplify the formulae for the luminosity distance, such as ``observational coordinates''~\cite{Ellis} or ``geodesic lightcone coordinates'' (GLC coordinates)~\cite{Fanizza:2013, Gasperini, Fleury}, (see also Appendix \ref{glc}). More recent approaches are~\cite{Hellaby} and~\cite{Miko} --- the latter considers the optical drift of various cosmological observables in a general spacetime.

The structure of this paper is the following. In Section 2 we look at the different notions of cosmological distances and the relations between them. In Subsection 2.1 we look at the different contributions to the redshift in a general spacetime. In Subsection 2.2 we look at the affine parameter distance. In Subsection 2.3 we look at the van Vleck determinant and obtain formulae for the luminosity and area distances in terms of it. In Subsection 2.4 we consider the Jacobi determinant and its relation to the van Vleck determinant. In Subsection 2.5 we remark on the limitations of the luminosity and area distances.  In Section 3 we consider conformal deformations of the spacetime metric. After introducing the transformation properties under conformal transformations of various quantities, we look at two examples --- FLRW in Subsection 3.2 and Conformally FLRW in Subsection 3.3, both of them conformally related to the Einstein static universe. We calculate the van Vleck determinant for the Einstein static universe and analyse the three different cases and obtain formulae for the luminosity and area distances in FLRW and CFLRW. We also consider the limit of small peculiar redshifts. In Subsection 3.4 we conjecture formulae for the luminosity and area distances in a general spacetime. We conclude in Section 4.

Throughout the paper we work with the convention $c=1$ and the metric is taken to have signature $(-1, 1, 1, 1)$.

\section{Luminosity distance and area distance}\label{S:luminosity}

The luminosity distance is, at its most fundamental, defined in terms of the energy flux at the observer, $F_o$, and the (absolute) luminosity  at the source, $L_s$, (the total output power). See for instance~\cite{Bonvin:2005,Visser:2003}, or any of many other sources.
\begin{equation}
F_o =  {L_s\over 4\pi \; d_L^2}.
\end{equation}
Measuring the photon energy flux $F_o$ is in principle straightforward.  In counterpoint, estimating the source luminosity $L_s$ is trickier, and model-dependent, but once this is somehow achieved, an estimate for the luminosity distance is
\begin{equation}
d_L  =   \sqrt{L_s\over4\pi \; F_o}. 
\end{equation}
To connect these definitions to specific properties of the spacetime, we start by noting that geometrically and kinematically one can argue that
\begin{equation}
F_o = {1\over (1+z)^2}\;  {L_s\over 4\pi (dA_o/d\Omega_s)}.
\end{equation}
Here $dA_o$ is the cross-sectional area, measured at the observer, of a spray of null geodesics which are, at the source, emitted into a solid angle $d\Omega_s$.\footnote{A ``null geodesic spray'' is simply a ``null geodesic congruence'' emitted from a specified point.}  (So the $dA_o/d\Omega_s$ factor is purely geometric in origin.) 

The two redshift factors are purely kinematical, due to 
two known effects: Individual photon energies are suppressed by a redshift factor $1/(1+z)$, whereas the arrival time between individual photons is stretched by a redshift factor $(1+z)$. 
The total redshift in turn depends on the peculiar velocities of both source and observer, on possible local gravitational inhomogeneities at both source and observer, and on the cosmological expansion of space between the source and observer. 

Then in any arbitrary spacetime, without any loss of generality
\begin{equation}
d_L  =   \sqrt{L_s\over4\pi \; F_o} =   (1+z)\;\sqrt{dA_o\over d\Omega_s} .
\end{equation}
(See for instance~\cite{Bonvin:2005}.) Up to this stage, this is actually a quite standard result. 
Now observationally, instead of measuring the photon energy flux it is actually easier to count the number of photons received by the observer, and not appreciably more difficult to estimate the number of photons emitted by the source over the total lifetime of the event, (or over some intrinsically defined interval, say from peak luminosity to half-maximum). Then in terms of integrated photon number flux, and photon number luminosity:
\begin{equation}
(F_\#)_o =  {(L_\#)_s\over 4\pi (dA_o/d\Omega_s)}.
\end{equation}
This eliminates the two explicit redshift factors, and so allows one to empirically define the ``area distance'' (see figure \ref{F:extended-observer}):
\begin{equation}
d_{area}  =   \sqrt{(L_\#)_s\over4\pi \; (F_\#)_o} =   \sqrt{dA_o\over d\Omega_s} .
\end{equation}
This area distance is arguably closer to the empirical observations than the more common luminosity distance.

\begin{figure}[h!]
\centering
\includegraphics[width=0.5\textwidth]{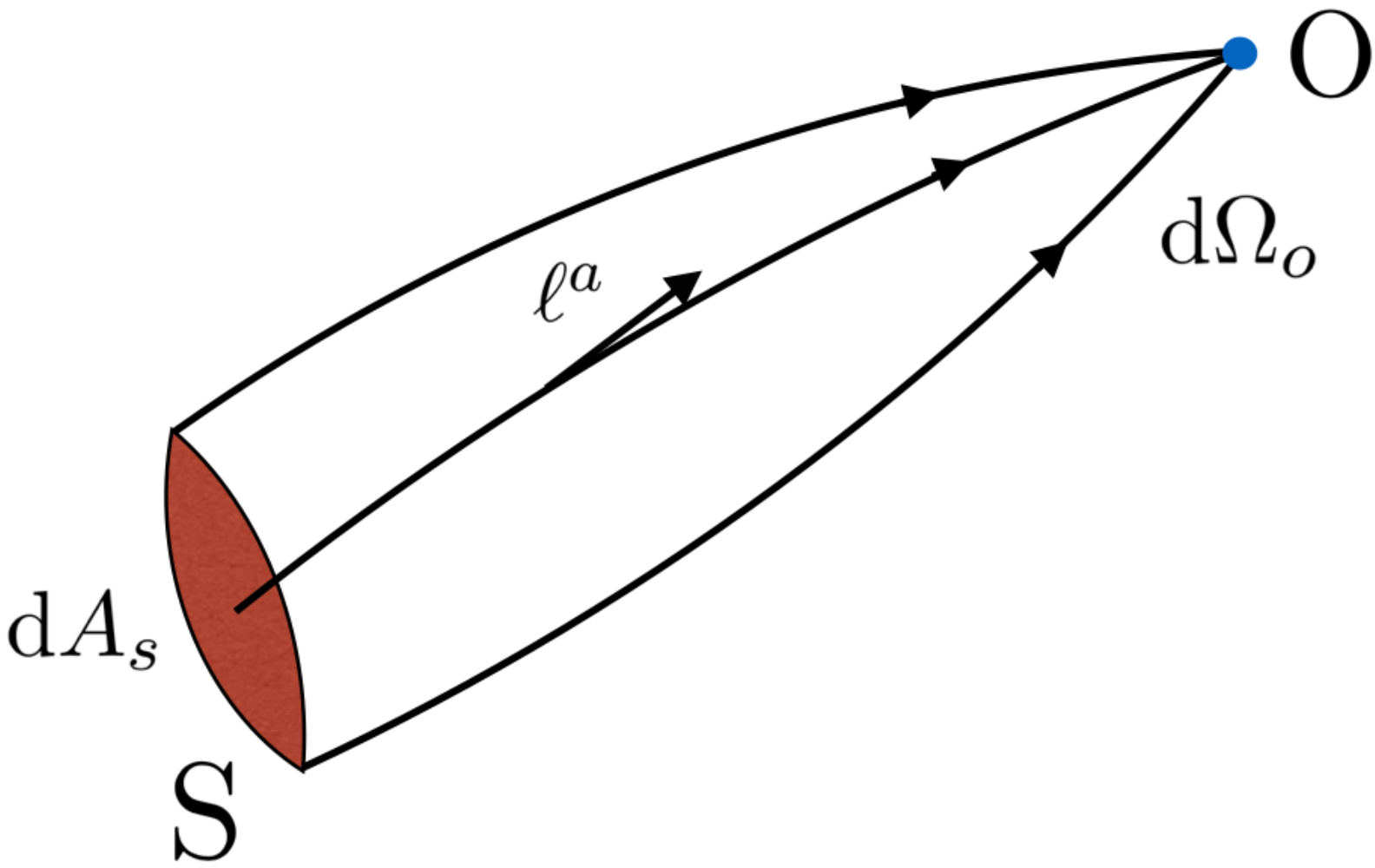}
\caption{\label{F:extended-source}
A congruence of light rays emitted at an extended source $S$ and received by a point observer $O$. The angular diameter distance between $S$ and $O$ is given by $d_{angular} = \sqrt{\mathrm{d}A_s/\mathrm{d}\Omega_o}$. Here $\ell^{a}$ is a tangent vector to a geodesic in the congruence. 
Under suitable technical conditions the Etherington distance duality relation yields $\sqrt{dA_o/d\Omega_s}  = (1+z) \;  \sqrt{dA_s/ d\Omega_o}$ .
}
\end{figure}

There is also a closely related notion of ``angular diameter distance", which can most easily be defined in terms of the physical size of the source and the angle subtended at the observer as
\begin{equation}
d_{angular} = \sqrt{dA_s\over d\Omega_o} .
\end{equation}
(We have seen both $d_{area}$ and $d_{angular}$ abbreviated as $d_A$; so we shall eschew the use of $d_A$ in thus article, and carefully distinguish area and angular diameter distances.)
The so-called ``Etherington distance duality'' relation, which is most commonly written as $d_L=(1+z)^2\, d_{angular}$,
becomes $d_{area}=(1+z)\, d_{angular}$ or even more explicitly (see figures \ref{F:extended-observer} and \ref{F:extended-source}):
\begin{equation}
 \sqrt{dA_o\over d\Omega_s}  = (1+z) \;  \sqrt{dA_s\over d\Omega_o} .
\end{equation}
The Etherington distance duality relation does depend on a number of assumptions.
Namely: (1) photon number is conserved, (2)
gravity is described by a metric theory,  (3) photons travel on unique null geodesics.
Exotic physics could in principle violate one or more of these assumptions, and observational tests of this
duality relation are an active area of research~\cite{Bassett:2003,Kunz:2004,Nair:2015}.
There are in addition many other notions of cosmological distance current in the literature~\cite{Hogg:1999}.
For current purpose we shall however focus on luminosity and area distances.

While luminosity distances are for historical reasons more traditional~\cite{Bonvin:2005,Sasaki:1987,Futamase:1989,Pyne:2003,Barausse:2005,Hui:2005,Fanizza:2013,Marozzi:2014,Yoo:2016,Umeh:2012,Umeh:2014,DiDio:2016,Visser:2003,Visser:2004,Cattoen:2007a,Cattoen:2007b,Cattoen:2008,Visser:2009,Vitagliano:2009,Visser:2015,Visser:2015b}, there are good physics reasons for working with area distances as well. (Of course one has not completely eliminated the redshift, just the manifestly obvious redshift factors --- we shall soon see that when trying to connect ${dA_o\over d\Omega_s}$ back to the evolution of the universe the cosmological contribution to the redshift will sneak back in.) From now on we shall work mainly with the luminosity distance keeping in mind that for each formula for the luminosity distance there exists an equivalent formula for the area distance which is obtained by removing the $1+z$ factor.

\subsection{{Redshift}}

To define the redshift it is convenient to consider a null geodesic (a photon trajectory), affinely parameterized by $\lambda$, and carefully distinguish the null tangent 4-vector, the null 4-wave-vector, and the null 4-momentum:
\begin{equation}
\ell^a = {d x^a\over d\lambda};  \qquad  
k^a = \tilde \omega \, \ell^a; \qquad 
p^a = \hbar \,k^a = \hbar \tilde \omega \, \ell^a = \tilde E \, \ell^a.
\end{equation}
Since the photon 4-momentum is, by definition, parallel transported along the null trajectory 
we have $\ell^a \nabla_a p^b=0$, (no external forces act on the photon, it is in free-flight). Since the tangent vector is chosen to be affine parameterized we have $\ell^a \nabla_a \ell^b=0$. Consequently
the scalar $\tilde E = \hbar \tilde  \omega$, (and hence $\tilde\omega$ itself), is constant along the null trajectory. 
However $\tilde E$ is not the locally measured energy, and $\tilde\omega$ is not the locally measured frequency.
In general for an observer of 4-velocity $V^a$ one has
\begin{equation}
E = -g_{ab} V^a p^b = \tilde E \; (-g_{ab} V^a \ell^b);  \qquad 
\omega = -g_{ab} V^a k^b = \tilde \omega \; (-g_{ab} V^a \ell^b).
\end{equation}
So $E$ and $\omega$ can change by purely geometric factors along the photon trajectory, as they should.

Now let the source have timelike 4-velocity $(V_s)^a$, and the observer have timelike 4-velocity $(V_o)^a$. Then in all generality one has the rigorous non-perturbative result that the total redshift is (see for example~\cite{Bonvin:2005,Visser:2015}):
\begin{equation}
1+z 
= {(g_{ab} \;p^a V^b)_s\over (g_{ab} \;p^a V^b)_o} 
= {(g_{ab} \;k^a V^b)_s\over (g_{ab} \;k^a V^b)_o} 
=  {(g_{ab} \;\ell^a V^b)_s\over (g_{ab} \;\ell^a V^b)_o}. 
\end{equation}
Note that the total redshift is purely geometrical, and by definition automatically frequency independent (achromatic). Let us now introduce two fiducial 4-velocities, $(W_s)^a$ and $(W_o)^a$, at the source and observer and write
\begin{equation}
1+z 
=  {(g_{ab} \;\ell^a W^b)_s\over (g_{ab} \;\ell^a W^b)_o}\;
  {(g_{ab} \;\ell^a V^b)_s\over (g_{ab} \;\ell^a W^b)_s}  \;
  {(g_{ab} \;\ell^a W^b)_o\over (g_{ab} \;\ell^a V^b)_o}. 
\end{equation}
These fiducial 4-velocities $W^a$ might represent, for instance, the local rest frame of the CMB, or the local rest frame of the Hubble flow, though we do not at this stage need to make any stringent assumptions along these lines; any fiducial 4-velocity would do. This now factorizes the total redshift into an overall cosmological/gravitational contribution, plus two peculiar velocity contributions. First, the factor
\begin{equation}
1+z_{*} =  {(g_{ab} \;\ell^a W^b)_s\over (g_{ab} \;\ell^a W^b)_o}
\end{equation}
represents the combined effects (as seen by fiducial observers) of cosmological expansion plus possible local variations in the gravitational field. 
Second, the factors
\begin{equation}
1+z_p =  {(g_{ab} \;\ell^a V^b)\over (g_{ab} \;\ell^a W^b)} = \gamma(1- \hat \ell\cdot \vec v)
\end{equation}
represent the effect of peculiar velocities of source/observer 4-velocities $V^a$ relative to the fiducial background $W^a$.  (In the absence of any choice of fiducial observer $W^a$ one cannot even begin to define the notion of ``peculiar velocity''.)  Here we have gone to Riemann normal coordinates at both source and observer, so $g_{ab}\to \eta_{ab}$, and have gone to the fiducial rest frame $W^a\to(1;\vec 0)$, with $V^a\to\gamma(1;\vec v^i)$ and $\ell^a\propto(1; \hat \ell^i)$  to cast the peculiar redshifts in the $ \gamma(1-\hat \ell\cdot \vec v)$ form.  {(Note that insofar as we are dealing with peculiar velocities we have not needed to normalize the null vector, in that $\ell^a\propto(1; \hat \ell^i)$ merely  defines a direction, and merely implies $\ell^a= \ell^0(1; \hat \ell^i)$. But $\ell^0$ drops out of the calculation.)} Overall we have (in all generality) the elegant factorized form
\begin{equation}
1+z = (1 + z_{*}) \; {[\gamma(1-\hat \ell\cdot \vec v)]_s\over[\gamma(1-\hat \ell\cdot \vec v)]_o} 
= (1+z_{*}) \; {1+z_{p,s}\over 1+z_{p,o}}= (1+z_{*}) \;(1+z_D) .
\end{equation}
This neatly splits the total redshift into cosmological/gravitational contributions plus peculiar motion contributions. This version of the redshift equation in principle allows for arbitrarily high peculiar velocities, and arbitrarily high cosmological/gravitational redshifts.\footnote{Often the peculiar velocities are known or assumed to be small, $|\vec v| \ll 1$, (we have set $c\to1$), in which case one has the simple perturbative result $1+z = (1 + z_{*}) \; \left(1-\hat \ell\cdot [\vec v_s-\vec v_o]+ O(v^2) \right)$. 
But such a low-peculiar-velocity approximation is by no means necessary.
Sometimes one sees the Doppler factor written as $\sqrt{(1+\hat\ell\cdot\vec v)/(1-\hat\ell\cdot\vec v)}$, but this is actually \emph{wrong} as it incorrectly ignores the transverse Doppler effect. (Though it does give the correct low-peculiar-velocity limit.)
}
(Note that peculiar velocities need not necessarily be intrinsically small. Ultimately it is an observational question as to just how small they are~\cite{Davis:2010}.)

\subsection{{Affine parameter distance}}

Consider the spray of affinely parameterized null geodesics, with null tangent 4-vector $\ell^a = dx^a/d\lambda$, emitted from the source into some solid angle $d\Omega_s$. While the null affine parameters are well defined (up to constant rescaling) along each individual null geodesic, there is \emph{a priori} no connection between the affine parameters on distinct null geodesics. 
Introduce such a connection between distinct null geodesics by enforcing (at the source and for  
all elements of the geodesic spray):
\begin{equation}
(g_{ab}\; \ell^a W^b)_s = -1.
\end{equation}
Going (temporarily) to Riemann normal coordinates at the source, $(g_s)_{ab}\to \eta_{ab}$, and going to the fiducial rest frame $(W_s)^a\to(1;\vec 0)$, this implies (at the source and for all elements of the  geodesic spray) that {we can set}
\begin{equation}
\ell^a =  (1; \hat \ell^i).
\end{equation}
(Though note that $\hat \ell^i$ will vary as one selects different elements of the geodesic spray; the geodesics in the geodesic spray are moving more-or-less in the same direction, they are not moving \emph{exactly} in the same direction.)
That is, the affine parameters for all the null geodesics of interest are normalized by asserting that, at the source, the affine parameter {along any of the elements of the null geodesic spray} equals the proper time of the fiducial observer {at the apex of that geodesic spray}. That is, the fiducial observer at the source sees an outgoing (partial) light cone, and using local flatness, uses his/her time coordinate to naturally fix the near-apex affine parameter on each of the null geodesics making up the (partial) light cone. With this normalization convention the affine parameter distance $\Delta \lambda = \lambda_o-\lambda_s$ equals the distance the photon \emph{would have travelled in Minkowski space} if the Riemann tensor were forced to zero.\footnote{With this normalization convention the cosmological/gravitational contribution to the redshift simplifies to 
$1+z_* = [(g_{ab} \;\ell^a W^b)_o]^{-1}$. This apparent simplification is less useful than one might at first imagine since one has to propagate the chosen normalization from source to observer.}
As we shall soon see the luminosity distance $d_L$ is proportional to affine parameter distance $\Delta \lambda= \lambda_o-\lambda_s$ modulated by redshift and focussing effects:
\begin{equation}
d_L = (1+z) \; (\lambda_o-\lambda_s) \; \Delta_{vV}^{-1/2}.
\end{equation}

\subsection{{The van Vleck determinant}}

The so-called van Vleck determinant has a wide and deep theoretical base in the general relativity, quantum mechanics, and quantum field theory communities. The van Vleck determinant is a general tool of wide applicability. In a general relativistic context see for instance~\cite{Visser:1992,Visser:1993,Visser:1995,Ottewill:2009,Chu:2011}, further afield see for instance~\cite{deBoer:1995,Jeon:2004,Barvinsky:2005,Neville:1982,vanVleck}. Among other things, in a general relativity context the van Vleck determinant encodes deviations from the inverse square law in the expansion of a null geodesic spray~\cite{Visser:1992,Visser:1993,Visser:1995}. 

A formal definition of the van Vleck determinant is given in ~\cite{Visser:1992,Visser:1993}. Here we explicitly evaluate the van Vleck determinant starting from equation (21.60) of reference~\cite{Visser:1995}:
\begin{equation}
{d \Delta_{vV}\over d\lambda} = \left({2\over\lambda} - \theta\right) \Delta_{vV}. \label{a3}
\end{equation}
Note the source is taken to be at $\lambda=0$. In flat Minkowski spacetime  one has $\theta = 2/\lambda$ and so $\Delta_{vV} = 1$. In a general setting the quantity $\left({2\over\lambda} - \theta\right)$ is the difference between what the null geodesic expansion would have been in Minkowski space and the actual null geodesic expansion in the curved spacetime of interest.

Integrate from some small $\epsilon$ to $\lambda$. We see
\begin{equation}
\ln \Delta_{vV}(\lambda) - \ln \Delta_{vV}(\epsilon) = 2 \ln \lambda - 2 \ln \epsilon- \int_\epsilon^\lambda \theta(\lambda') d \lambda' .
\end{equation} 
Then
\begin{equation}
 \Delta_{vV}(\lambda) = \lambda^2 \exp\left(  \ln \Delta_{vV}(\epsilon) -  2 \ln \epsilon- \int_\epsilon^\lambda \theta(\lambda') d \lambda' \right).
\end{equation} 
But as $\epsilon\to 0$ we have $\Delta_{vV}(\epsilon)\to 1$ as a byproduct of local flatness.
So a more accurate and precise statement is this:
\begin{equation}
 \Delta_{vV}(\lambda) = \lambda^2 \exp\left(  \lim_{\epsilon\to0} \left[ - \int_\epsilon^\lambda \theta(\lambda') d \lambda' -2 \ln\epsilon\right] \right). \label{a1}
\end{equation} 
If we now shift the source from $\lambda=0$ to $\lambda_s$ then we have
\begin{equation}
{d \Delta_{vV}\over d\lambda} = \left({2\over\lambda-\lambda_s} - \theta\right) \Delta_{vV}, \label{a3}
\end{equation}
and
\begin{equation}
\theta(\lambda) = {2\over\lambda-\lambda_s} + O(1),
\end{equation}
and all that happens to the formalism is this:
\begin{equation}
 \Delta_{vV}(\lambda) = (\lambda-\lambda_s)^2 \exp\left(  \lim_{\epsilon\to0} \left[ - \int_{\lambda_s+\epsilon}^\lambda \theta(\lambda') d \lambda' - 2\ln\epsilon\right] \right), \label{a2}
\end{equation} 
To match this to the previous analysis, note the standard result that:
\begin{equation}
\theta =   {1\over dA/d\Omega_s} {d(dA/d\Omega_s)\over d\lambda}.
\end{equation}
See for instance the texts by Hawking and Ellis~\cite{Hell} or Wald~\cite{Wald}; the point is that the null expansion is the relative change in cross sectional area $A$ of a bundle of null geodesics (of fixed solid angle $d\Omega_s$) emitted from the source.
Then we see
\begin{equation}
\int_{\lambda_s+\epsilon}^\lambda \theta(\lambda') d \lambda'
=
\int_{\lambda_s+\epsilon}^\lambda {1\over dA/d\Omega_s} {d(dA/d\Omega_s)\over d\lambda'} d \lambda' =
\ln\left(dA_\lambda/d\Omega_s\right) - \ln\left(dA_{\lambda_s+\epsilon}/d\Omega_s\right).
\end{equation}
Thence
\begin{equation}
\Delta_{vV}(\lambda) = (\lambda-\lambda_s)^2 \exp\left(  \lim_{\epsilon\to0} \left[ - \ln\left(dA_\lambda/d\Omega_s\right) + \ln\left(dA_{\lambda_s+\epsilon}/d\Omega_s\right)
- 2\ln\epsilon\right] \right).
\end{equation} 
So
\begin{equation}
\Delta_{vV}(\lambda) = {(\lambda-\lambda_s)^2\over \left(dA_\lambda/d\Omega_s\right)}
 \exp\left(  \lim_{\epsilon\to0} \left[ \ln\left(dA_{\lambda_s+\epsilon}/d\Omega_s\right)
- 2\ln\epsilon\right] \right).
\end{equation} 
But at small distances from the source local flatness implies
\begin{equation}
\left(dA_{\lambda_s+\epsilon}/d\Omega_s\right) = \epsilon^2 + O(\epsilon^3).
\end{equation}
So 
\begin{equation}
 \lim_{\epsilon\to0} \left[ \ln\left(dA_{\lambda_s+\epsilon}/d\Omega\right)- 2\ln\epsilon\right] =0,
\end{equation}
and we have
\begin{equation}
\Delta_{vV}(\lambda) = {(\lambda-\lambda_s)^2\over \left(dA_\lambda/d\Omega_s\right)}.
\end{equation}
In particular at the observer
\begin{equation}
\Delta_{vV}(\lambda_o) = {(\lambda_o-\lambda_s)^2\over \left(dA_o/d\Omega_s\right)}.
\end{equation}
So the luminosity distance is
\begin{equation}
d_L = (1+z) \sqrt{dA_o\over d\Omega_s} 
= (1+z) \; (\lambda_o-\lambda_s) \; \Delta_{vV}^{-1/2},
\end{equation}
as required. 
So as anticipated, in any arbitrary spacetime we have 
\begin{equation}
d_L = (1+z) \; (\lambda_o-\lambda_s)\; \Delta_{vV}^{-1/2};
\qquad
d_{area} = (\lambda_o-\lambda_s)\; \Delta_{vV}^{-1/2}.
\end{equation}
The nice feature here is that the affine parameter distance $\Delta\lambda = \lambda_o-\lambda_s$ is directly related to the length of the photon trajectory, and that the van Vleck determinant $\Delta_{vV}^{-1/2}$ is directly related to focussing/defocussing effects (and trivializes to 1 in flat Minkowski space). If desired, one can relate this back to the flux by writing (in any arbitrary spacetime)
\begin{equation}
F_o =  {1\over(1+z)^2} \;  {L_s  \; \Delta_{vV} \over 4\pi(\lambda_o-\lambda_s)^2}.
\end{equation}
This makes manifest the fact that the van Vleck determinant characterizes deviations from the inverse square law~\cite{Visser:1992,Visser:1993,Visser:1995}.

\subsection{The Jacobi determinant}  

The  $dA_o/ d\Omega_s$ factor, and so implicitly the van Vleck determinant, can be related to the determinant of the so called Jacobi map~\cite{Schneider, Bonvin:2005}. The Jacobi map relates changes in the null geodesic tangent vector at the source to transverse shifts in the photon trajectories. While the full Jacobi map contains more information than its determinant --- information about the shear and vorticity of the congruence~\cite{Fanizza} --- it is only its determinant which is important for the analysis of the luminosity and area distances. Here we relate the van Vleck determinant directly to the Jacobi determinant without considering the full Jacobi map.

For a simplified presentation, consider the null geodesic spray emitted from the source in a solid angle $d\Omega_s$ aimed in the general direction of the observer. 
In the immediate vicinity of the observer, the surface of constant $\Delta\lambda = \lambda_o-\lambda_s$ is generically a spacelike 2-surface of solid angle $d\Omega_s$, (modulo possible focussing singularities to be briefly discussed below). On that 2-surface set up angular 2-coordinates $\xi^i=(\theta,\phi)$ so that the 2-surface is characterized by $x^a(\xi^i)=x^a(\theta,\phi)$. Then the induced 2-metric on that 2-surface is simply
\begin{equation}
g_{ij}(\xi^k)  = g_{ab}(x^c(\xi^i)) 
\left(\partial x^a\over\partial \xi^i\right) 
\left(\partial x^b\over\partial \xi^j\right).
\end{equation}

Now consider 
\begin{equation}
{dA_o\over d\Omega_s}  
= {\sqrt{\det(g_{ij})} \; d\Omega_s \over d\Omega_s} 
= \sqrt{\det(g_{ij})} 
=
 \sqrt{\det\left\{g_{ab} \left(\partial x^a\over\partial \xi^i\right) 
\left(\partial x^b\over\partial \xi^j\right)\right\}}.
\end{equation}
Choose a vierbein/tetrad $e^A{}_b$ at the observer so that $g_{ab} = \eta_{AB} \, e^A{}_a\, e^B{}_b$ then
\begin{equation}
{dA_o\over d\Omega_s}  
=
\sqrt{\det\left\{ \eta_{AB} \; e^A{}_a\, e^B{}_b \; \left(\partial x^a\over\partial \xi^i\right) 
\left(\partial x^b\over\partial \xi^j\right)\right\}}.
\label{E:2-pdet}
\end{equation}
(In an Appendix A we relate this to the not particularly well known but increasingly useful notion of ``pseudo-determinant''.)

To finally relate this to the Jacobi matrix and Jacobi determinant, consider the 2-plane, at the observer, that is tangent to the 2-surface of constant $\Delta\lambda = \lambda_o-\lambda_s$. Set up Cartesian coordinates $x_\perp^i$ on this tangent plane, and consider the mapping $\xi^i \to x^i_\perp(\xi^i)$ from the solid angle coordinates to the 2-tangent plane.  The Jacobi matrix is then
\begin{equation}
J^i{}_j = {\partial x^i_\perp \over \partial\xi^j},
\end{equation}
and we have
\begin{equation}
{dA_o\over d\Omega_s}  
= \left|\det(J)\right| = \left|\det\left({\partial x^i_\perp \over \partial\xi^j}\right)\right|.
\end{equation}
Then in terms of the Jacobi determinant we have the following results
\begin{equation}
d_L = (1+z) \sqrt{\left|\det(J)\right|}; \qquad d_{area} = \sqrt{\left|\det(J)\right|}.
\end{equation}
(The Jacobi determinant does not separate the inverse-square and focussing contributions, they are simply lumped together.)
Comparing with the calculation for the van~Vleck determinant we have
\begin{equation}
\left|\det(J)\right| = (\lambda_o-\lambda_s)^2 \; \Delta_{vV}^{-1}.
\end{equation}
Here we see the  inverse-square and focussing contributions cleanly separated. 

\subsection{{Limitations of the luminosity and area distances}}

There is an implicit and fundamental limitation to using luminosity distance, and hence also the area distance, that should be kept in mind:
At any conjugate point on the null geodesic~\cite{Hell} one has both $\det(J)=0$ and $\Delta_{vV} = \infty$,
so the luminosity distance is \emph{zero} at any null geodesic focal point.
(In fact the luminosity distance is \emph{zero} at any null caustic.)
As a practical matter, this suggests that any background galaxy with image sitting on top of, (or sufficiently near to), an Einstein arc is likely to have a  luminosity distance that is grossly 
misleading. A foreground galaxy, between us and the lensing object, would not be anywhere near as problematic. (In counterpoint, while we have never observationally seen net negative gravitational masses, these would also lead to caustics and imply unusual behaviour for the luminosity distance~\cite{natural}.) This is not, \emph{per se}, a fatal objection --- but it is something to keep in mind. This is also an opportunity --- supernovae in host galaxies near an Einstein arc might in principle allow one to test aspects of otherwise unaccessible fundamental physics --- but we will defer any such considerations for the future.

Even in a completely smooth FLRW universe the luminosity distance can sometimes be misleading: Consider a 
$k=+1$ FLRW universe (spatially a hyper-sphere), or a $k=0$ FLRW universe with topological identifications (spatially a hyper-torus), or a $k=-1$ FLRW universe with topological identifications (spatially the hyperbolic plane modded out by a M\"obius group). In all these cases with non-simply connected topology there is a finite spatial size to the universe and sufficiently large distances become small again. As a practical matter, this only causes significant problems for static universes, with eternally radiating non-evolving stars and galaxies --- otherwise we could look at the ``age'' of the object to decide if it was near or far.
There is at best extremely limited evidence for these extremely large-scale universe-sized problems. In contrast medium scale lensing and null caustics are potentially an issue.

\section{{Conformal deformations of the spacetime metric}}
\subsection{Generalities}

The behaviour of null geodesics under conformal deformations of the metric implies specific transformation properties for the luminosity distance. Under a conformal deformation $g_{ab} = \exp(2\Phi) \hat g_{ab}$ the paths of the null geodesics are unaffected, 
though the affine parameters are non-trivially related by~\cite{Hell,Wald,Visser:1995}
\begin{equation}
d\lambda =  \exp(2\Phi) d\hat\lambda; \qquad \ell^a = \exp(-2\Phi) \hat \ell^a.
\end{equation}
Trivially one has
\begin{equation}
{dA_o\over d\Omega_s}   = \exp(2\Phi_o)\; {d\hat A_o\over d\Omega_s}. 
\end{equation}
More subtle is the fact that
\begin{equation}
g_{ab} \ell^a V^b =  (\exp(2\Phi) \hat g_{ab}) (\exp(-2\Phi) \hat \ell^a) (\exp(-\Phi) \hat V^a)
=  \exp(-\Phi) \; (\hat g_{ab} \hat \ell^a \hat V^b).
\end{equation}
This implies that the redshift transforms as
\begin{equation}
1+z =  \exp(\Phi_o-\Phi_s)\; (1+\hat z).
\end{equation}
Hence the luminosity distance transforms as
\begin{equation}
d_L =  (1+z) \; \sqrt{dA_o\over d\Omega_s}    
= \exp(2\Phi_o-\Phi_s) \; (1+\hat z) \; \sqrt{d\hat A_o\over d\Omega_s} = \exp(2\Phi_o-\Phi_s) \; \hat d_L,
\end{equation}
while the area distance transforms as
\begin{equation}
d_{area} =   \sqrt{dA_o\over d\Omega_s}    
= \exp(\Phi_o)  \; \sqrt{d\hat A_o\over d\Omega_s} = \exp(\Phi_o) \; \hat d_{area}.
\end{equation}
That is
\begin{equation}
d_L = \exp(2\Phi_o-\Phi_s) \; \hat d_L; \qquad d_{area} = \exp(\Phi_o) \; \hat d_{area}. 
\end{equation}
In applications one would typically take $g_{ab}$ to be the physical spacetime metric while $\hat g_{ab}$ would be some (not directly physical) conformal deformation of spacetime useful for calculational purposes.
Observe that the Jacobi matrix has the simple transformation law
\begin{equation}
J^i{}_j = \exp(\Phi_o) \; \hat J^i{}_j; \qquad \det(J) =  \exp(2\Phi_o)\; \det(\hat J);
\end{equation}
but that the behaviour of the affine parameter distance (and van Vleck determinant) is more subtle. Indeed 
\begin{equation}
\Delta \hat \lambda = \hat\lambda_o - \hat\lambda_s = \int_s^o \exp[-2\Phi(\lambda')] \; d \lambda'.
\end{equation}
The fact that affine parameter distance (and van Vleck determinant) does not simply rescale under conformal deformations is actually potentially useful, not a hindrance --- it implies that the van Vleck determinant might actually simplify much more radically under conformal deformations.

\subsection{{Example: FLRW cosmologies}}

A particularly simple example of conformal deformations of the spacetime metric comes from considering FLRW cosmologies. Consider the usual FLRW cosmology with spacetime metric\footnote{Here $k\in\{-1,0,+1\}$ while $r$ is dimensionless and $a$ has units of distance. Note that the possibility that $k\neq0$ is currently undergoing somewhat of a resurgence in popularity~\cite{Ratra}.}:
\begin{equation}
ds_{FLRW}^2 =   
- dt^2 +a(t)^2 \left\{ {dr^2\over1-kr^2} + r^2(d\theta^2+\sin^2\theta\;d\phi^2) \right\}.
\end{equation}
Introduce a new time parameter (conformal time) by defining $dt/a(t) = d\eta/a_*$, where $a_*$ is some convenient arbitrary but fixed constant, (needed for $k=\pm1$ in order to satisfy dimensional analysis),
and then subsequently reparameterize  $a(t)\to a(\eta)$ to write:
\begin{equation}
ds_{FLRW}^2 =  g_{ab} dx^a dx^b = \left[a(\eta)\over a_*\right]^2 \left\{ - d\eta^2 + a_*^2 \left[ {dr^2\over1-kr^2} + r^2(d\theta^2+\sin^2\theta\;d\phi^2)\right] \right\}.
\end{equation}
Now the term in braces 
\begin{equation}
ds_{E}^2 =  \hat g_{ab} dx^a dx^b = - d\eta^2 + a_*^2 \left[{dr^2\over1-kr^2} + r^2(d\theta^2+\sin^2\theta\;d\phi^2) \right],
\end{equation}
is just the spacetime metric for the Einstein static universe, which we shall use as a computational aid; the real physical universe not being static. 
Then 
\begin{equation}
d_{L,FLRW} =  \left(a_o^2\over a_s a_*\right) \; d_{L,E}.
\end{equation}
The point is that $d_{L,E}$, the luminosity distance in the unphysical but computationally useful Einstein static universe, is extremely easy to calculate. For a source placed at $r=0$, and observer located at $r$, the flux is simply
\begin{equation}
F = {1\over (1+z_E)^2} \;   {L\over4\pi (a_* r)^2}.
\end{equation}
Here $z_E$ the redshift in the Einstein static universe, depends only on the peculiar velocities (there are no local inhomogeneities or cosmological effects) so
\begin{equation}
d_{L,E} = (1+z_E) \; a_* r = \left(1+z_{p,s}\over1+z_{p,o}\right) a_* r.
\end{equation} 
But to determine $r$ in terms of conformal time we note that for a null geodesic in the Einstein static universe 
\begin{equation}
d\eta = a_*  \; {dr\over\sqrt{1-kr^2}},
\end{equation}
so we have\footnote{For $k=0$ interpret this as a limit: $\lim_{k\to0} \{\arcsin(\sqrt{k} x)/\sqrt{k}\} = x$; whereas for $k=-1$ one simply has 
$ \{\arcsin(\sqrt{-1} x)/\sqrt{-1}\} =  \{\arcsin(i x)/i \} = \arcsinh(x)$.}
\begin{equation}
\Delta \eta = \eta_o-\eta_s =  a_*  \int_0^r {d\tilde r\over\sqrt{1-k\tilde r^2}} 
= a_* {\arcsin\left(\sqrt{k}r\right)\over\sqrt{k}},
\end{equation}
which implies\footnote{For $k=0$ interpret this as a limit: $\lim_{k\to0} \{\sin(\sqrt{k} x)/\sqrt{k}\} = x$; whereas for $k=-1$ one simply has 
$ \{\sin(\sqrt{-1} x)/\sqrt{-1}\} =  \{\sin(i x)/i \} =\sinh(x)$.}
\begin{equation}
r =  {\sin\left(\sqrt{k}\Delta \eta/a_*\right)\over\sqrt{k}}.
\end{equation}
Now in the Einstein static universe we have
\begin{equation}
{dA_o\over\d\Omega_s} =  (a_* r)^2 =  {a_*^2 \sin^2\left(\sqrt{k}\Delta \eta/a_*\right)\over{k}},
\end{equation}
and furthermore the affine null parameter is simply the conformal time $\lambda_E = \eta$,
so we can extract the van Vleck determinant as
\begin{equation}
\Delta_{vV,E} = { k (\Delta \eta /a_*)^2 \over \sin^2\left(\sqrt{k}\Delta \eta/a_*\right)}.
\end{equation}
Then for $k=0$ (flat spatial slices) $\Delta_{vV,E} \equiv1$, so there is no deviation from the inverse square law. For $k=-1$ (hyperbolic spatial slices) $\Delta_{vV,E} =  {(\Delta \eta /a_*)^2 \over \sinh^2\left(\Delta \eta/a_*\right)}< 1$. That is, one has defocussing induced suppression of the inverse square law. 
Finally, for $k=1$ (hyperspherical spatial slices) $\Delta_{vV,E} =  {(\Delta \eta /a_*)^2 \over \sin^2\left(\Delta \eta/a_*\right)}>1$, one has focussing induced enhancement of the inverse square law.

Now going to the physical FLRW spacetime one has 
\begin{equation}
d_{L,FLRW} =  \left(a_o^2\over a_s\right) \; \left(1+z_{p,s}\over1+z_{p,o}\right) \; r 
= a_o \; (1+z_*) \; \left(1+z_{p,s}\over1+z_{p,o}\right) \; r .
\end{equation}
Here $z_*$ is the cosmological contribution to the FLRW redshift (in this simple model there are no local inhomogeneities). 
Furthermore
\begin{equation}
d_{L,FLRW} =  a_o \; (1+z_*) \; \left(1+z_{p,s}\over1+z_{p,o}\right) \;
 {\sin\left(\sqrt{k}\Delta \eta/a_*\right)\over\sqrt{k}}.
\end{equation}
Thence we have the fully non-perturbative result
\begin{equation}
d_{L,FLRW} =  a_o \; (1+z_*) \; \left(1+z_{p,s}\over1+z_{p,o}\right) \; 
{\sin\left[\sqrt{k}\int_s^o {dt\over a(t)}\right]\over\sqrt{k}}.
\end{equation}
Though somewhat formal, this does have the theoretical advantage of nicely and cleanly separating out the various distinct physical contributions to the luminosity distance in a general FLRW spacetime.
For the area distance we have the fully non-perturbative result
\begin{equation}
d_{area,FLRW} =  a_o \; 
{\sin\left[\sqrt{k}\int_s^o {dt\over a(t)}\right]\over\sqrt{k}}.
\end{equation}
While \emph{explicit} occurrences of redshift have been eliminated, there will still be \emph{implicit} dependence on redshift coming from any attempt at converting $\int_s^o {dt\over a(t)}$ to a power-series in cosmological redshift. See for instance~\cite{Visser:2003}.
Indeed the goal of any ``cosmographic'' analyses is to invert $a(\eta)$ to find $\eta(a)$ and use $a_o/a_s = 1+z_*$ to rewrite the conformal time as a function of the cosmological contribution to the redshift 
\begin{equation}
\Delta \eta = a_*  \; f(1+z_*);  \qquad  \hbox{with normalization} \qquad f(1)=0.
\end{equation}
Indeed we observe that $f(1+z_*)$ is simply another way of encoding the in-principle-arbitrary function $a(\eta)$.
In terms of this newly defined function $f(1+z_*)$ we see
\begin{equation}
d_{L,FLRW} =  
a_o\; (1+z_*) \; \left(1+z_{p,s}\over1+z_{p,o}\right) 
\;  {\sin\left[\sqrt{k}\,f(1+z_*) \right]\over\sqrt{k}}.
\end{equation}
Therefore
\begin{equation}
d_{L,FLRW} =  
a_o\; (1+z)  
\;  {\sin\left[\sqrt{k}\,f\left( [1+ z] \left[ 1+z_{p,o}\over1+z_{p,s}\right] \right) \right]\over\sqrt{k}}. \label{f1}
\end{equation}
This gives (in any FLRW spacetime) a non-perturbative formulation for luminosity distance as a function of total redshift plus the peculiar velocity induced redshifts. The function $f(1+z)$ can be written in the standard manner in terms of the Hubble parameter, scale factor, deceleration, jerk,  snap and higher order parameters. See for instance~\cite{Visser:2003}.

These formulae clearly show the full redshift dependance of the luminosity and area distances.
They provide non-perturbative information on the way redshift comes in; something that is in principle a little deeper than immediately resorting to finitely truncated power series. By rewriting the cosmological redshift in terms of total redshift and peculiar redshifts we have arranged to have at least one directly measurable quantity $z$ in the final answer; and have at least some control over how the peculiar velocity contributions to the redshift contribute to the luminosity distance.

Specifically for small peculiar redshifts we see
\begin{eqnarray}
d_{L,FLRW} &=&
a_o\; (1+z)  \Bigg\{
\;  {\sin\left[\sqrt{k}\,f\left( 1+ z  \right) \right]\over\sqrt{k}}
\nonumber\\ && 
 \quad + {\cos\left[\sqrt{k}\,f\left( 1+ z  \right) \right] \; (1+z) \; f'(1+z) [z_{p,o}-z_{p,s}]} 
 + O([\delta z]^2) \Bigg\},\quad
\end{eqnarray}

This formula is exact in terms of the underlying FLRW cosmology, 
but approximate in terms of peculiar redshifts.

\subsection{{Example: CFLRW cosmologies}}

A \emph{conformal} FLRW spacetime (CFLRW spacetime) is simply any spacetime that is conformal to a FLRW spacetime, but with an arbitrary possibly position-dependent conformal factor~\cite{Visser:2015}. This in turn implies that any CFLRW spacetime is conformal to an Einstein static spacetime~\cite{Visser:2015}:
\begin{equation}
ds_{CFLRW}^2 =  g_{ab} \, dx^a dx^b = \left[a(x)\over a_*\right]^2 \left\{ - d\eta^2 + a_*^2 \left[ {dr^2\over1-kr^2} + r^2(d\theta^2+\sin^2\theta\;d\phi^2)\right] \right\}.
\end{equation}
The physical reason that CFLRW spacetimes are important is because of the observed smoothness of the CMB: In view of the fact that the CMB is smooth to $O(10^{-5})$,  the conformal mode is the only deviation from FLRW that has the slightest chance of becoming non-perturbatively large (between last scattering an the current epoch) without grossly distorting the CMB~\cite{Visser:2015}.  The mathematical reason that CFLRW spacetimes are important is because (from the point of view of luminosity distance, area distance,  and photon propagation) they are almost as easy to analyze as the FLRW spacetimes. Although the CFLRW cosmologies are relatively easy to analyze mathematically, to date only relatively crude bounds have been put on the conformal factor~\cite{Visser:2015}.
(More complicated deviations from FLRW are generally treated perturbatively, though there is ongoing debates as to the significance of non-perturbative effects. See for instance~\cite{Buchert:2007,Rasanen:2006,Clarkson:2011,Meures:2011,Enqvist:2009,Rasanen:2009}.)

The analysis of photon propagation in the Einstein static (unphysical) reference spacetime carries through (completely unchanged) as in the previous subsection, and going to the physical CFLRW spacetime one has 
\begin{equation}
d_{L,CFLRW} =  a_o \; (1+z_*) \; \left(1+z_{p,s}\over1+z_{p,o}\right) \;  {\sin\left(\sqrt{k}\Delta \eta/a_*\right)\over\sqrt{k}}.
\end{equation}
Here $a_o$ can now be both time and space dependent, while $z_*$ is the cosmological/local gravity contribution to the CFLRW redshift. In this model there \emph{are} local conformal inhomogeneities, so $1+z_* = a_o/a_s$. One could for instance factor the scale factor $a(x) = \bar a(\eta) \; a_{local}(x)$ into a contribution $\bar a(\eta)$ characterizing the overall change in volume of the spatial slices, and a second contribution $a_{local}(x)$ characterizing volume-preserving conformal distortions of the spatial slices~\cite{Visser:2015}. Then the cosmological and local contributions to the redshift further factorize 
\begin{equation}
(1+z_*) = {a_o\over a_s} = {\bar a_o\over\bar a_s} \; {(a_{local})_o\over (a_{local})_s} = (1+\bar z_*) (1+z_{local}). 
\end{equation}
Collecting all this one has the non-perturbative result
\begin{equation}
d_{L,CFLRW} =  
\bar a_o\; (a_{local})_o \; (1+\bar z_*) \;(1+z_{local})\; \left(1+z_{p,s}\over1+z_{p,o}\right) 
\;  {\sin\left(\sqrt{k}\,\Delta \eta/a_*\right)\over\sqrt{k}}.
\end{equation}
Though definitely rather formal, this does have the theoretical advantage of nicely and cleanly separating out the various distinct physical contributions to the luminosity distance in a general CFLRW spacetime. Note in particular,  that the factor $a_0 = \bar a_o\; (a_{local})_o $ is common to all the objects one might look at. So there is no real loss of generality to simply absorbing this into one's definition of distance and asserting 
\begin{equation}
d_{L,CFLRW} \propto
(1+\bar z_*) \;(1+z_{local})\; \left(1+z_{p,s}\over1+z_{p,o}\right) 
\;  {\sin\left(\sqrt{k}\,\Delta \eta/a_*\right)\over\sqrt{k}}.
\end{equation}
This is equivalent to considering a luminosity \emph{modulus} instead of a luminosity \emph{distance}, $\mu_{L,CFLRW} = \ln(d_{L,CFLRW} /a_*)$, and agreeing to ignore a common offset. 

In counterpoint, adopting a modified ``cosmographic'' analyses (similar to that for the FLRW spacetimes) one can invert $\bar a(\eta)$ to find $\eta(\bar a)$, and use $\bar a_o/\bar a_s = 1+\bar z_*$ to rewrite the conformal time as a function of the cosmological contribution to the redshift:\begin{equation}
\Delta \eta = a_*  \; f(1+\bar z_*).
\end{equation}
This implies
\begin{equation}
d_{L,CFLRW} =  
a_o\; (1+\bar z_*) \;(1+z_{local})\; \left(1+z_{p,s}\over1+z_{p,o}\right) 
\;  {\sin\left[\sqrt{k}\,f(1+\bar z_*) \right]\over\sqrt{k}}.
\end{equation}
Therefore
\begin{equation}
d_{L,CFLRW} =  
a_o\; (1+ z)  
\;  {\sin\left[\sqrt{k}\,f\left( \left[1+ z\over1+z_{local}\right] \left[ 1+z_{p,o}\over1+z_{p,s}\right] \right) \right]\over\sqrt{k}}. \label{f3}
\end{equation}
(Note we now have a contribution from $z_{local}$, the extra redshift contribution due to local deformations of the conformal factor, as well as from the peculiar velocity contributions.)
This quite formal result seems to be as far as one can go without making some approximations and resorting to perturbation theory. 

Remember, $z$ is easy to measure; whereas measuring or estimating $z_{local}$ and $z_p$ is significantly more tricky.

For small peculiar redshifts, and small $z_{local}$, we now see
\begin{eqnarray}
d_{L,CFLRW} &=&
a_o\; (1+z)  \Bigg\{
\;  {\sin\left[\sqrt{k}\,f\left( 1+ z  \right) \right]\over\sqrt{k}}
\nonumber\\ && 
 \quad + {\cos\left[\sqrt{k}\,f\left( 1+ z  \right) \right] \; (1+z) \; f'(1+z) [z_{p,o}-z_{p,s} - z_{local}]} 
 \nonumber\\ && 
 \quad + O([\delta z]^2) \Bigg\},
\end{eqnarray}

This formula is exact in terms of the underlying CFLRW cosmology, 
but approximate in terms of peculiar redshifts and $z_{local}$.

\subsection{General spacetimes}

In view of the results presented above we can argue that non-perturbatively the best we can hope for in any completely general spacetime is that
\begin{equation}
d_{L} =  
a_o\; (1+z) \; F(\bar z_*);
\qquad
d_{area} =  
a_o \; F(\bar z_*);
\qquad
F(0)=0;
\end{equation}
for some function $F(\bar z_*)$ of the cosmological contribution to the total redshift
\begin{equation}
1+ z =  (1+\bar z_*) \;(1+z_{local})\; \left(1+z_{p,s}\over1+z_{p,o}\right).
\end{equation}
Then
\begin{equation}
d_{L} =  
a_o\;(1+z) \; F\left( \left[1+ z\over1+z_{local}\right] \left[ 1+z_{p,o}\over1+z_{p,s}\right] -1\right);
\end{equation}

A quite common approximation is to simply set
\begin{equation}
z_{local} \approx z_{p,s} \approx z_{p,o}\approx 0;  \qquad  z \approx \bar z_*;
\end{equation}
so that
\begin{equation}
d_{L} \approx  a_o\; (1+\bar z_*) \; F(\bar z_*) \approx a_o\;(1+z)\;F(z) ;  
\qquad d_{area} = a_o\; F(\bar z_*) \approx a_o\; F(z). 
\end{equation}
A better approximation would be to at least retain first-order terms in non-cosmological contributions to the redshift, $z_{local}$, $z_{p,s}$, and  $z_{p,o}$, a task to which we will turn in future work~\cite{future}.
For now let us just point out that to first order
\begin{equation}
d_L =  
a_o \; (1+z) \; \left\{ F(z) - (1+z)\,F'(z) [z_{local}-z_{p,o}+z_{p,s}] + O([\delta z]^2)  \right\};
\end{equation}

\section{Conclusion}
In this article we have derived expressions for the luminosity and area distances in terms of the redshift, the affine parameter distance and the van Vleck determinant which hold for an arbitrary spacetime. They are given by:
\begin{equation}
d_L =  (1+z) \; (\lambda_o-\lambda_s) \; \Delta_{vV}^{-1/2};
\qquad
d_{area} =  (\lambda_o-\lambda_s) \; \Delta_{vV}^{-1/2}.
\end{equation}

We have also shown that in all generality the Jacobi determinant is related to the van~Vleck determinant by:
\begin{equation}
\left|\det(J)\right| = (\lambda_o-\lambda_s)^2 \; \Delta_{vV}^{-1}.
\end{equation}

The van Vleck determinant explicitly captures deviations from the inverse square law. In flat spacetime, where the inverse square law holds exactly, we have $\Delta_{vV} =1$. In an arbitrary spacetime one might have $\Delta_{vV} >1$, corresponding to geodesic focussing, or $\Delta_{vV} < 1$, corresponding to geodesic defocussing. So the van Vleck determinant allows you to clearly separate the inverse square law contribution from the focussing/defocussing contribution; information which is obscured in the Jacobi determinant.

We have also derived explicit expressions for $d_L$ in arbitrary FLRW and Conformally FLRW Universes --- they are given by \eqref{f1},  \eqref{f3}. Similar expressions hold for $d_{area}$ with the $1+z$ factor removed.

While both $d_L$ and $d_{area}$ have limitations due to the potential existence of conjugate points along the null geodesic congruence and the potentially nontrivial topology of the universe, they allow you to test certain aspects of the geometry of the universe and, when complemented by redshift drift, allow you to test dynamics too.

\appendix
\section{{Pseudodeterminants}}

It is potentially of interest to re-write the formula (\ref{E:2-pdet}) for $dA_o/d\Omega_s$ in terms of the increasingly commonly used notion of pseudo-determinant:
\begin{equation}
{dA_o\over d\Omega_s}  
=
\sqrt{\det\left\{ \eta_{AB} \; e^A{}_a\, e^B{}_b \; \left(\partial x^a\over\partial \xi^i\right) 
\left(\partial x^b\over\partial \xi^j\right)\right\}}
= \left|\pdet\left( e^A{}_a \; {\partial x^a \over \partial \xi^i}\right)\right|.
\end{equation}
Here pdet, the pseudo-determinant, is simply the product over nonzero eigenvalues. Note that $\pdet(X)$ is sometimes denoted $\det'(X)$, especially in the QFT literature. Furthermore $\pdet(AB)$ is not necessarily equal to $\pdet(A)\,\pdet(B)$, some care is required in the analysis.  Observe that the $4\times2$ matrix $ e^A{}_a \; {(\partial x^a/ \partial \xi^i)}$ has at least two eigen-zeros. One way of seeing the need for the pseudo-determinant  is to note that the set of null geodesic tangent vectors forms a 2-sphere, the celestial sphere at the source.

To establish this result, argue as follows: Set $M^A{}_i =  e^A{}_a \; {\partial x^a \over \partial \xi^i}$, this is a $4\times2$ matrix. Now the vectors $M^a{}_i \, d\xi^i =  e^A{}_a \; {\partial x^a \over \partial \xi^i}\; d\xi^i$ are always spacelike, (since one is working on the spacelike 2-surface at constant $\lambda_o$). This implies the existence of a singular value decomposition of the form
\begin{equation}
M^A{}_i = L^A{}_B  \;\;
\left[\begin{array}{cc}0&0\\0&0\\ \Lambda_1  & 0 \\ 0& \Lambda_2 \end{array} \right]^B_{\;\;j} \;\; R^j{}_i.
\end{equation}
Here $L$ is a $4\times4$ Lorentz transformation and $R$ is a $2\times2$ rotation.
Then (suppressing indices on the matrix multiplications) 
\begin{equation}
M^T \eta M 
=   R^T  \left[\begin{array}{cccc}0&0&\Lambda_1  & 0 \\0&0& 0 &\Lambda_2 \end{array} \right]   
     L^T \; \eta \;  L  \left[\begin{array}{cc}0&0\\0&0\\ \Lambda_1  & 0 \\ 0 &\Lambda_2 \end{array} \right] R.
\end{equation}
But by the definition of a Lorentz transformation $L^T \eta L = \eta$, so
\begin{equation}
M^T \eta M 
=   R^T  \left[\begin{array}{cccc}0&0&\Lambda_1  & 0 \\0&0& 0 &\Lambda_2 \end{array} \right]   \eta \left[\begin{array}{cc}0&0\\0&0\\ \Lambda_1  & 0 \\ 0 &\Lambda_2 \end{array} \right] R.
\end{equation}
Multiplying out the inner three matrices
\begin{equation}
M^T \eta M 
=   R^T  \left[\begin{array}{cc}\Lambda_1^2  & 0 \\ 0 &\Lambda_2^2 \end{array} \right]   R.
\end{equation}
But then
\begin{equation}
\det(M^T \eta M) =   \Lambda_1^2 \; \Lambda_2^2,
\end{equation}
so that
\begin{equation}
\sqrt{\det(M^T \eta M)} =   \left|\Lambda_1 \; \Lambda_2 \right|.
\end{equation}
That is
\begin{equation}
\sqrt{\det(M^T \eta M)} =   \left|\pdet(M)\right|,
\end{equation}
which completes the proof.

\section{{Geodesic light cone coordinates}}

\label{glc}

Geodesic light-cone (GLC) coordinates are a special type of coordinate system centred at a particular observer or source~\cite{Gasperini, Fleury}. They are set up in such a way that the formulas for the luminosity and area distances simplify considerably~\cite{Fanizza:2013}. In certain cases, the formula obtained for the luminosity distance can be transformed to a more familiar gauge like the Poisson gauge~\cite{Dayan}. Here we outline the derivations of the expressions for the Jacobi and van Vleck determinants and area and luminosity distances in the GLC gauge.

In geodesic light-cone coordinates (\emph{aka} GLC gauge) the spacetime metric is:
\begin{equation}
ds^2  = \Upsilon^2 dw^2 -2 \Upsilon dw d\tau + \gamma_{ij}(d\xi^i - U^i dw)(d\xi^j-U^j dw); \qquad   \{i,j\}\in\{1,2\}.
\end{equation}
If we order the coordinates as $x^a=(\tau,w,\xi^i)$ then one has:
\begin{equation}
g_{ab} =  \left[ \begin{array}{c|c||c} 
                0&-\Upsilon&0\\
                \hline
                -\Upsilon & \Upsilon^2 + \gamma_{ij} U^i U^j& - U_j\\
                \hline\hline
                0&-U_i &\gamma_{ij}
                \end{array}\right]
\qquad\qquad
g^{ab} =  \left[ \begin{array}{c|c||c} 
                -1&-\Upsilon^{-1}&U^j/\Upsilon\\
                \hline
                -\Upsilon^{-1} & 0 & 0\\
                \hline\hline
                -U^i/\Upsilon & 0  &\gamma^{ij}
                \end{array}\right],
\end{equation}
and
\begin{equation}
\det(g_{ab}) = \Upsilon^2 \det(\gamma_{ij}); \qquad\qquad
\det(g^{ab}) = \Upsilon^{-2} \det(\gamma^{ij}).
\end{equation}

Here are some elementary observations:
\begin{itemize}
\item 
The vector field $V^a = -g^{ab} \nabla_b \tau = (1, \Upsilon^{-1}; U^i/\Upsilon)^a$ satisfies $g_{ab}V^aV^b=-1$, and so is an affinely parameterized timelike geodesic vector field. One has
\begin{equation}
0= \nabla_c (g_{ab}V^aV^b) = \nabla_c (g^{ab}V_aV_b) = 2 g^{ab} V_a \nabla_c V_b = 
2 g^{ab} V_a \nabla_b V_c = 2 V^b \nabla_b V^c = 0.
\end{equation}
\item
The vector field $\ell^a = -g^{ab} \nabla_b  w = (\Upsilon^{-1},0;0)^a$ satisfies $g_{ab}\ell^a\ell^b=0$, and so is an affinely parameterized null geodesic vector field. One has
\begin{equation}
0= \nabla_c (g_{ab}\ell^a\ell^b) = \nabla_c (g^{ab}\ell_a\ell_b) = 2 g^{ab} \ell_a \nabla_c \ell_b = 
2 g^{ab} \ell_a \nabla_b \ell_c = 2 \ell^b \nabla_b \ell^c = 0.
\end{equation}
\item
Note that on the null geodesics ${d\tau\over d\lambda} = \Upsilon^{-1}$, so the affine parameter is $d\lambda = \Upsilon d\tau$. 
\item
The ``centre'' of the spacetime is parameterized by $w$, and given by the timelike trajectory $x^a(w)= (\tau_*(w),w;\xi^i)^a$ with tangent vector $t^a(w) = (\tau'_*(w), 1; 0)^a$. 

\item
In the vicinity of the ``centre'' local flatness implies
\begin{equation}
ds^2 \approx dw^2 -2 dwd\tau + [\tau-\tau_*(w)]^2 \hat \gamma_{ij} d\xi^i d\xi^j,
\end{equation}
where $\hat ds^2 = \hat \gamma_{ij} d\xi^i d\xi^j$ is some (arbitrary) coordinate representation of the metric for a unit-radius constant-curvature 2-sphere. 

\item
When properly normalized, (using local flatness), the 4-velocity of the ``centre'' is 
\begin{equation}
U^a = {(\tau'_*(w), 1; 0)^a\over 2 \tau'_*(w)-1}.
\end{equation}

\item
Note that $\ell^a = -g^{ab} \nabla_b  w = (\Upsilon^{-1},0;0)^a$, so the null geodesics follow lines of constant $w$ and constant $\xi^i$.
\item 
A bundle of null geodesics emitted from, or impinging on, the centre subtends a solid angle $d\Omega=\sqrt{\det{\hat \gamma} } \;d^2\xi$.
\item 
Away from the centre, a bundle of ingoing or outgoing null geodesics has cross sectional area $dA = \sqrt{\det{ \gamma} } \;d^2\xi$.
\item
The ``centre'' of the spacetime can either be set at the observer or at the source; the construction of GLC coordinates being largely symmetric under interchange of source and observer. Setting the centre 
at the observer maximizes the extent to which these GLC coordinates align with the so-called ``observer coordinates''. Setting the centre 
at the source maximizes the extent to which these GLC coordinates align with the rest of the present article. Let us therefore choose to put the centre at the source. 

\end{itemize}

Then we have:
\begin{equation}
{dA_0\over d\Omega_s} = {\sqrt{\det(\gamma_{\,o}) }\over \sqrt{\det(\hat \gamma_{\,s}) }}.
\end{equation}
This implies:
\begin{equation}
|\det (J)| = {\sqrt{\det(\gamma_{\,o}) }\over \sqrt{\det(\hat \gamma_{\,s})}};
\qquad
\Delta_{vV} (\lambda_o) = (\lambda_o - \lambda_s)^2 {\sqrt{\det(\hat \gamma_{\,s}) }\over \sqrt{\det(\gamma_{\,o})}};
\end{equation}

\begin{equation}
d_{area} = {\sqrt[4]{\det(\gamma_{\,o}) }
\over \sqrt[4]{\det(\hat \gamma_{\,s}) }};
\qquad
d_L = (1+z) \; {\sqrt[4]{\det( \gamma_{\,o}) }
\over \sqrt[4]{\det(\hat \gamma_{\,s}) }}.
\end{equation}

So in GLC coordinates the area and luminosity distance simplify considerably (they factorise into source and observer contributions) while the complexity is hidden in the way the coordinates are set up.

As a consistency check, consider the 4-current density:
\begin{equation}
j^a =  {\ell^a\over\sqrt{\det(\gamma)}} = {(1,0;0)^a\over\sqrt{\det(g)}}.
\end{equation}
Then
\begin{equation}
\nabla_a j^a 
= {1\over\sqrt{g}} \partial_a ( \sqrt{g} j^a )
= {1\over\sqrt{g}} \partial_a \left( \Upsilon \sqrt{\gamma} \; {(\Upsilon^{-1},0;0)^a\over\sqrt{\gamma}} \right)
= {1\over\sqrt{g}} \partial_a \left(1,0;0\right)^a = 0.
\end{equation}
So this current is conserved in the general relativistic sense.

\enlargethispage{10pt}
This implies that the flux is
\begin{equation}
F_{o}  
= 
{1\over(1+z)^2} \;{L\over4\pi} \; {d\Omega_{s}\over d A_{o}}
=
{1\over(1+z)^2} \;{L\over4\pi} \;
 {\sqrt{\det(\hat \gamma)_{s}}\over \sqrt{\det(\gamma)_{o}}},
\end{equation}
which again yields the same simple factorized formulae for area distance and luminosity distance.

\section*{Acknowledgements}
Matteo Viel was partially supported  by the INFN PD51 INDARK grant.\\
Matt Visser was supported by the Marsden Fund, which is administered by the Royal Society of New Zealand.


\end{document}